\documentclass[prl,showpacs,preprintnumbers,amsmath,amssymb,letterpaper,twocolumn]{revtex4}

\usepackage{graphicx}
\usepackage{dcolumn}
\usepackage{bm}
\usepackage{epstopdf}
\usepackage{latexsym}
\usepackage{epsfig}
\usepackage{verbatim}
\usepackage{hyperref}

\newcommand{\be}{\begin{equation}}
\newcommand{\ee}{\end{equation}}
\newcommand{\bea}{\begin{eqnarray}}
\newcommand{\eea}{\end{eqnarray}}

\newcommand{\eref}[1]{Eq.~(\ref{#1})}

\begin{document}

\title{Measurement of Conduction Electron Polarization Via the
Pairing Resonance}

\author{Y.M. Xiong, P.W. Adams}
\affiliation{Department of Physics and Astronomy, Louisiana State
University, Baton Rouge,
Louisiana 70803, USA}
\author{G. Catelani}
\affiliation{Department of Physics and Astronomy, Rutgers University,
Piscataway, New Jersey 08854, USA}

\date{\today}

\begin{abstract}
We show that the pairing resonance in the Pauli-limited normal state
of ultra-thin superconducting Al films provides a spin-resolved probe
of conduction electron polarization in thin magnetic films. A
superconductor-insulator-ferromagnet tunneling junction is used to
measure the density of states in supercritical parallel magnetic
fields that are well beyond the Clogston-Chandresekhar limit, thus
greatly extending the field range of tunneling density of states
technique.  The applicability and limitations of using the pairing
resonance as a spin probe are discussed.

\end{abstract}


\maketitle

The possibility of incorporating spin degrees of freedom into
electronic technologies has led to an explosion in research
into mechanisms of spin polarization of conduction currents in
semiconducting and metallic systems \cite{Wolf2001,Prinz1995}. In
addition to being technologically important, spin polarization in
conducting systems remains a fundamentally interesting many-body
problem.   Clearly, an accurate determination of the conduction
electron polarization $P$, particularly in thin-film magnetic
structures, is crucial both from the point of view of
``spintronic'' device development and basic research. Unfortunately, however,
there are few direct probes of $P$.  The three most successful
techniques, listed in historical
order, have been Zeeman-split superconducting tunneling density of
states spectroscopy (SCTDoS) \cite{TM1,TM2}, spin-resolved photoemission spectroscopy (SRPES) \cite{Johnson1997}, and point
contact Andreev reflection (PCAR) \cite{Buhrman1998,Soulen1998}.
Each of these techniques has its own unique advantages and
limitations. SCTDoS has outstanding resolution $\sim10$ $\mu$V, and
is compatible with thin-film geometries, but is limited to a narrow
range of magnetic fields and is not easily implemented with bulk
samples \cite{TM2}.  SRPES has a relatively low resolution of $\sim$10 mV, and does not discriminate well between itinerant and localized bands.  PCAR can be used on bulk samples and thick
films, but it requires very low impedance point contacts, and is also
incompatible with magnetic fields \cite{Soulen1998,Buhrman2007}.  In
this Letter we introduce an extension of SCTDoS technique that
exploits the spin structure of the Pauli-limited normal state (PLNS)
pairing resonance \cite{Aleiner1997,Butko1999} to measure electron
polarization in magnetic fields well above the superconducting
critical field.   We show that the technique not only greatly
extends the range of fields over which the polarization can be
measured, but it can be much less sensitive to field misalignment than
SCTDoS.

Tedrow and Meservey \cite{TM2} pioneered the use of superconducting
spin-resolved tunneling to directly measure the electron
polarization in magnetic films.  The technique utilizes a planar
junction geometry consisting of a
superconductor-oxide-ferromagnet (SC-Ox-FM) sandwich in which the
superconductor counter-electrode is purposely made very thin. The
tunnel junction is then cooled below the superconducting transition
temperature and a carefully aligned magnetic field is applied
parallel to the junction plane.   If the SC film thickness $t$ is
much less than the superconducting coherence length $\xi$, then the
Meissner response to the applied field is suppressed, and the
critical field of the SC counter-electrode is entirely Zeeman
mediated \cite{Fulde1973}.  If the spin-orbit scattering rate of the
SC is small, then spin is a good quantum number, and the spin
rotation symmetry of the SC can be exploited to provide a
spin-resolved probe.  In the Tedrow and Meservey technique the
polarization is measured in the superconducting phase by applying a
subcritical parallel magnetic field to the tunnel junction at low
temperature, $T\ll T_c$.  This induces a Zeeman splitting of the SC
quasiparticle DoS spectrum, and the BCS coherence peaks get split
into spin-up and spin-down bands by the parallel field \cite{TM3}.
As will be discussed below, the relative heights of these peaks
give a direct measure of the electron polarization in the FM.  Here
we show that polarization can, in fact, be measured in fields that
are several times larger than the parallel critical of the SC by
using the PLNS pairing resonance in appropriately designed tunnel
junctions.

The SC-Ox-FM tunnel junctions were formed by first depositing a thin
Al counter-electrode via e-beam deposition of 99.999\% Al stock onto
fire polished glass microscope slides held at 84~K. The depositions
were made at a rate of $\sim0.1$~nm/s in a typical vacuum with
pressure $<3\times10^{-7}$ Torr.  After deposition, the
counter-electrode was exposed to the atmosphere for 10-20 minutes in
order to allow a thin native oxide layer to form. Then a 45~\AA\
thick FM film was deposited onto the counter-electrode with the
oxide serving as the tunneling barrier. In this study the FM was
either CNi$_3$ or CCo$_3$, where the e-beam depositions were made
from arc-melted buttons.  The counter-electrode thicknesses were chosen
so that their in-plane sheet resistance was
$\sim$1-2~k$\Omega$/sq. This corresponded to thicknesses that were typically
22-24~\AA.  The low temperature parallel critical fields of the
counter-electrodes were $\sim$6.5~T. The junction area was about
1~mm$\times$1~mm, while the junction resistance ranged from 15-100~k$\Omega$
depending on exposure time and other factors.  Only
junctions with resistances much greater than that of the films were
used.  Measurements of resistance and tunneling were carried out on
an Oxford dilution refrigerator using a standard ac four-probe
technique. Magnetic fields of up to 9~T were applied using a
superconducting solenoid. A mechanical rotator was employed to
orient the sample \textit{in situ} with a precision of
$\sim0.1^{\circ}$.

\begin{figure}
\begin{flushleft}
\includegraphics[width=.44\textwidth]{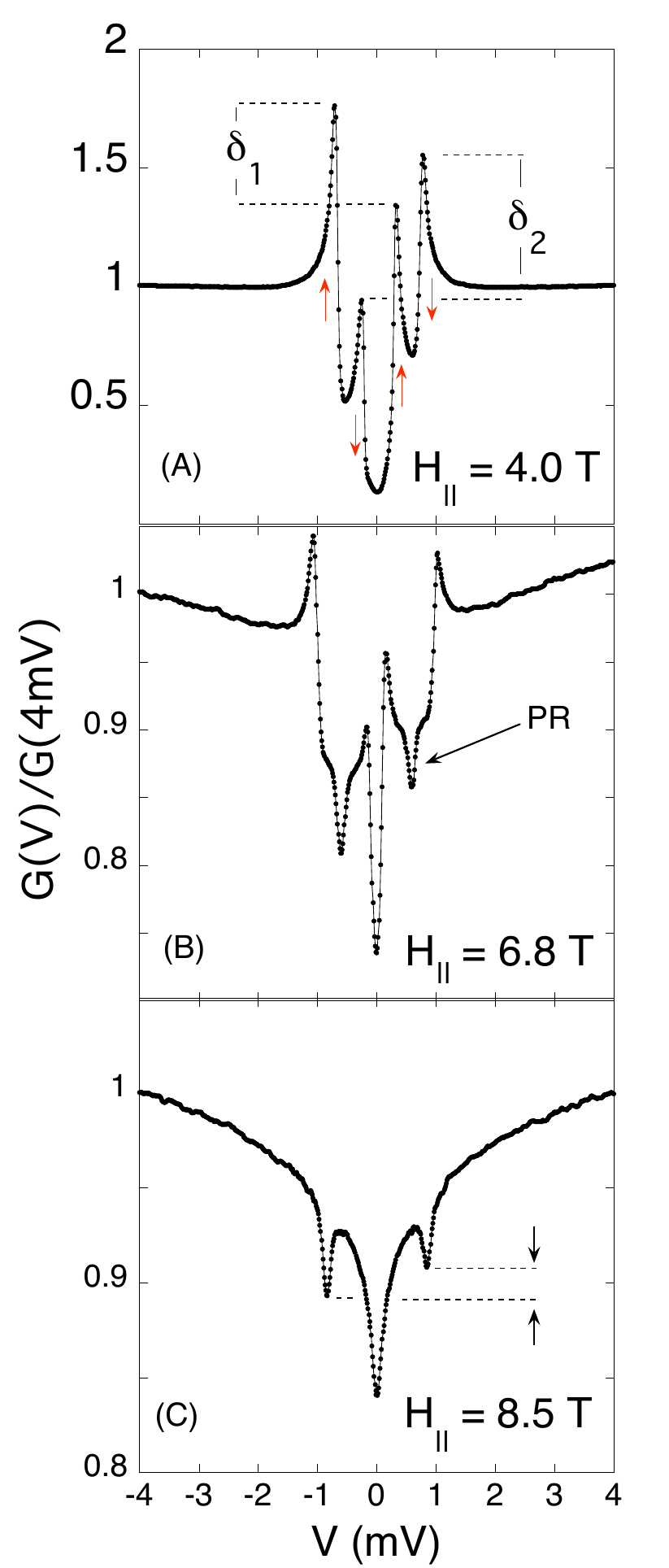}\end{flushleft}
\caption{\label{DoS}Evolution of the tunneling conductance of a
Al-AlO$_x$-CNi$3$ tunnel junction as the parallel critical field
transition is crossed at 70~mK.  A: superconducting phase showing an
asymmetric Zeeman-split DoS spectrum.  The arrows denote the spin
assignment of the coherence peaks. B: DoS spectrum at the parallel
critical field transition in which the normal-state pairing
resonance (PR) coexists with superconducting coherence peaks. C:
Pauli-limited normal state in which only the pairing resonance and
the zero bias anomaly remain.  Note that the positive and negative
resonances have different magnitudes.}
\end{figure}

In the upper panel of Fig.~\ref{DoS} we plot the 70~mK tunneling
conductance of a Al-AlO$_x$-CNi$_3$ tunnel junction in a
sub-critical parallel magnetic field of 4~T.  The Zeeman splitting
of the BCS DoS spectrum is clearly evident, where we have labeled
the spin moment associated with each peak.  Normally the respective
spin peaks would be identical on either side of the Fermi energy.  The
asymmetry arises from the unequal spin populations in the
ferromagnetic CNi$_3$.  This is expected since the magnetization
properties of the CNi$_3$ and CCo$_3$ are very similar to their
elemental counterparts \cite{CNi3}.  If one assumes that spin is conserved in
the tunneling processes, then tunneling currents from the spin-up
(down) bands in the Al will only tunnel into corresponding spin-up
(down) bands in the ferromagnet.  Tedrow and Meservey exploited this
conservation property to extract the polarization of a variety of
transition metal FM films \cite{TM2},
\be
P=\left|\frac{\delta_{1}-\delta_{2}}{\delta_{1}+\delta_{2}}\right|
\label{TM}
\ee
where the peak height differences $\delta_{1,2}$ are defined in
Fig.~\ref{DoS}A and
the corresponding polarization of the CNi$_3$ is $\sim18\%$.

The zero-field gap energy for the Al counter-electrode used in
Fig.~\ref{DoS} was determined to be $\Delta_o\sim0.49$~meV by fits to
the superconducting DoS spectrum.  Because the thickness of the Al
counter-electrode is much less than the superconducting coherence
length $\xi\sim 150$~\AA, it undergoes a first-order parallel critical
field transition at  the Clogston-Chandrasekhar critical field
\cite{Clogston1962,Chandrasekhar1962}
\begin{equation}
H^{CC}_{c}=\frac{\Delta_o\sqrt{1+G^0}}{\sqrt{2} \mu_B}
\label{Clogston}
\end{equation}
where $\mu_B$ is the Bohr magneton and $G^0$ is the
anti-symmetric Landau parameter \cite{CWA}. In panel (B) of
Fig.~\ref{DoS} we show the DoS spectrum at the critical field.  This
spectrum is in the coexistence region between the superconducting
phase and the Pauli-limited normal state.  The new feature that
appears at the transition is a manifestation of the pairing
resonance (PR).  Finally, in panel (C) we show a normal-state
spectrum where the BCS coherence peaks have been
extinguished.  The remaining structure consists of a broad,
symmetric, background with two small satellite resonances on either
side of V = 0.  The background feature is often referred to as the
zero bias anomaly and is a well documented electron-electron
interaction effect \cite{Altshuler1987,Imry1982}.  In contrast, the
satellite features represent incoherent Cooper pairing.  The fact
that the resonance dips have unequal magnitude in Fig.~\ref{DoS}
indicates that they are spin specific.  As we show below, they can be
used to extract FM polarization in fields well beyond the
Clogston-Chandrasekahr critical field of \eref{Clogston}.

\begin{figure}
\includegraphics[width=.46\textwidth]{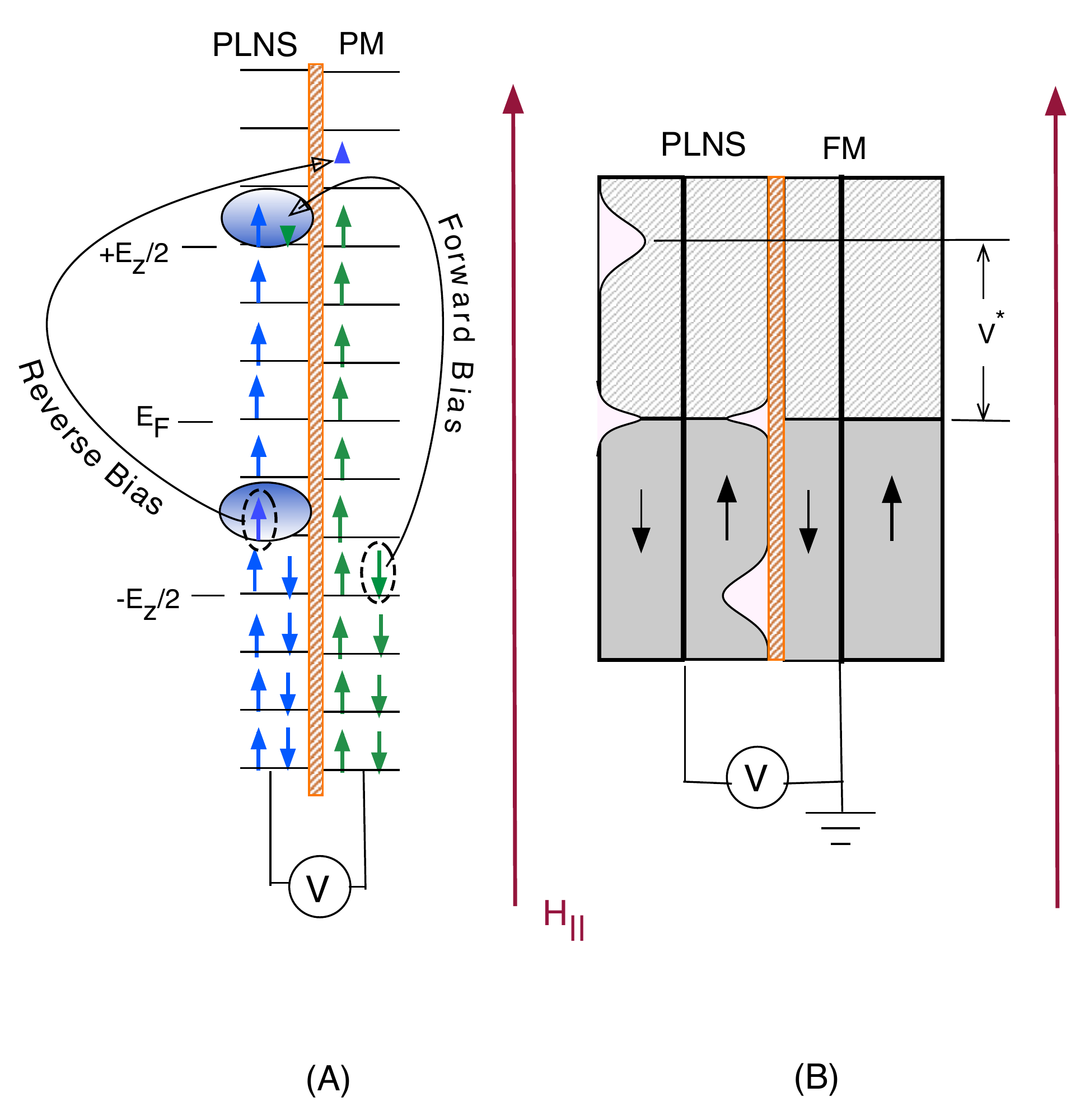}
\caption{\label{PRgraphic}A: Schematic of the spin structure of the
pairing resonance. B: DoS profiles of a tunnel junction comprised of
an Al film in the Pauli-limited normal state (PLNS) on one side and
a ferromagnetic film on the other.  Note the depletion of states in
the PLNS due to the pairing resonance and the zero bias anomaly.}
\end{figure}

For the purposes of measuring polarization the most important
property of the PR is its spin structure.  In Fig.~\ref{PRgraphic}
we present a graphic representation of the resonance as it is
observed in tunneling into a paramagnetic (PM) metal film.  Since
the PM has no preferred spin direction, the resonance is symmetric
about V = 0.  As is depicted in Fig.~\ref{PRgraphic}(A), when a
sufficient forward-bias voltage is reached, spin-down electrons in
the PM can tunnel across to form doubly occupied levels close to the top
of the spin-up band in the PLNS.  These spin-singlet states can then
mix with the unoccupied states in the near vicinity to form an
evanescent Cooper pair \cite{Aleiner1997}.  This effectively
produces a small depletion of spin-down quasiparticle states due to
the fact that they have been consumed by the resonance.  By
particle-hole symmetry there is a similar depletion of spin-up
states at the reverse-bias voltage needed for spin-up electrons lying
just above
the PLNS doubly occupied sites to tunnel over to the top of
the spin-up band in the PM.    The precise energy of these
resonances is field dependent,
\be
eV^* = \frac{1}{2}\left(E_z +\sqrt{E_z^2-\Delta_0^2}  \right) \ ,
\ee
where $E_z = (2\mu_B H)/(1+G^0)$ is the Zeeman energy renormalized by $G^0$.

\begin{figure}
\begin{flushleft}
\includegraphics[width=.48\textwidth]{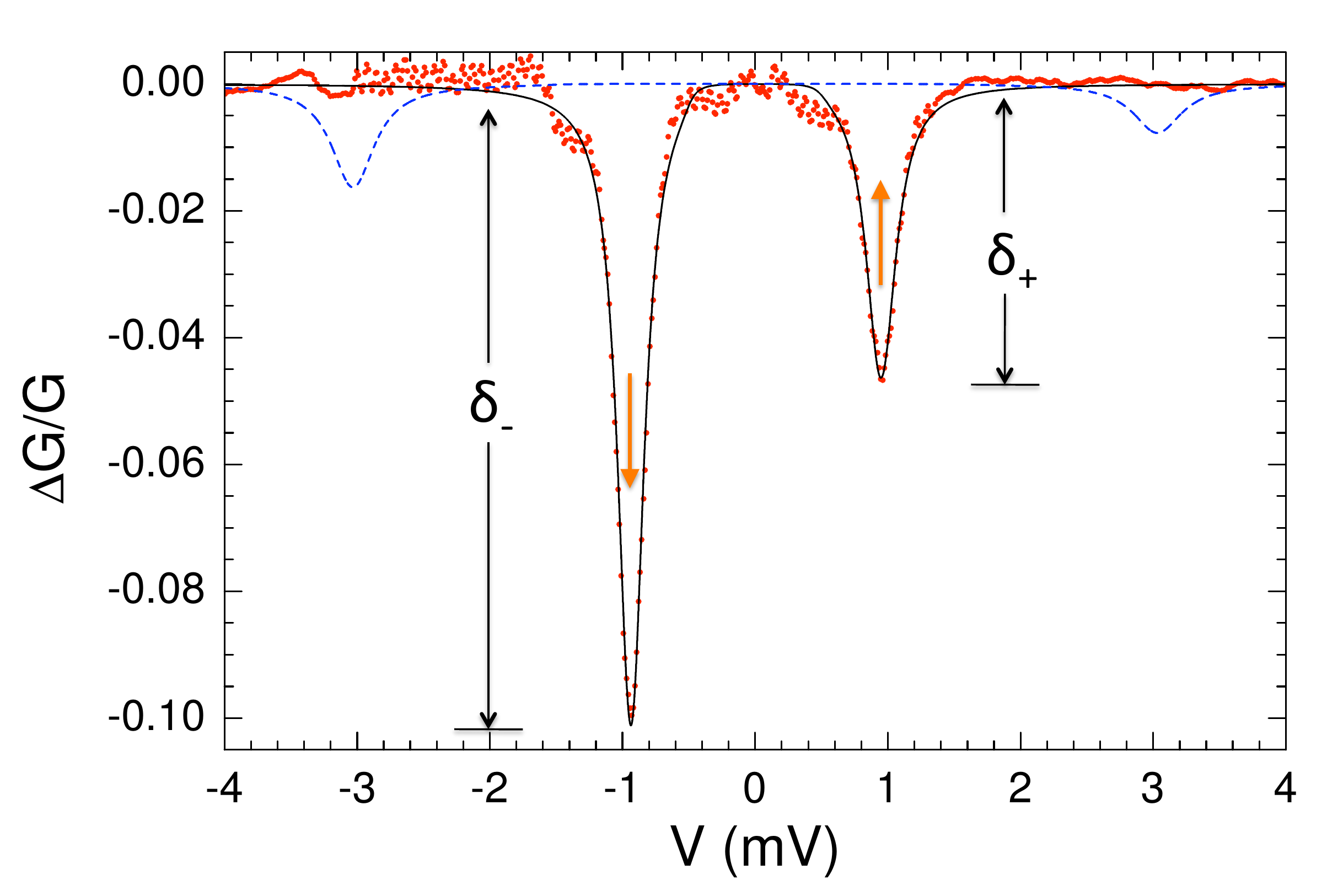}\end{flushleft}
\caption{\label{PR}The red symbols are tunneling spectra taken at
70~mK in a parallel field of 6.7~T, where the zero bias anomaly
background has been subtracted off.  The orange arrows denote the
spin assignment of the occupied and unoccupied resonances.  The
solid black line represents a best fit to the resonance curve.  The dashed blue
line is the predicted resonance profile at 18~T, extrapolating from the fitting
parameters obtained at 6.7~T.}
\end{figure}

The zero bias anomaly background in Fig.~\ref{DoS}(C) is independent
of magnetic field and varies as $\ln V$ for $V \gtrsim k_B T/e$.  It
can easily be subtracted from the data in order to isolate the
resonances as shown in Fig.~\ref{PR}.  The red dots are data taken
on a Al-AlO$_x$-CCo$_3$ tunnel junction at 70~mK in a 6.7~T parallel
magnetic field.  The arrows depicted the spin assignments of the
resonances and $\delta_{\pm}$ refer to their respective amplitudes.
The solid black line is a best fit to the resonance profile using a
procedure and formalism described elsewhere \cite{Catelani2009}.
As was the case for the superconducting phase, we only need the
relative magnitudes of the positive and negative bias resonances in
order to determine the polarization.  If we take the PLNS density of
states to be $N^s$, then well away from the PR spin rotation
symmetry requires $N_{\uparrow}^s=N_{\downarrow}^s=N^s/2$.  On
resonance, however, there will be a depletion of one spin component
at positive bias and the other at negative bias,
\be
N_{+,-}^s=N_{\downarrow,\uparrow}^s+N_{\uparrow,\downarrow}^s(1-\epsilon)
\label{Ns}
\ee
where $0\le\epsilon\le1$ represents the strength of
the resonance.  In general $\epsilon$ depends upon a number of
factors, including magnetic field, temperature, spin-orbit
scattering rate, and the dimensionless normal-state transport
conductance, $g$ \cite{Catelani2009,Aleiner1997}.  Spin conservation
forbids intra-spin-band tunneling, so if we assume that the
ferromagnet has a majority spin density $N_{\uparrow}^f$ and a
minority spin density $N_{\downarrow}^f$, then the tunneling
conductance is simply proportional to the product of respective
spin-specific DoS on either side of the tunnel junction as is
depicted in Fig.~\ref{PRgraphic}(B).  The magnitudes of the
positive- and negative-bias resonance features are
\begin{align}
\begin{split}
\delta_{+,-} &=
[N_{\downarrow,\uparrow}^sN_{\downarrow,\uparrow}^f+N_{\uparrow,\downarrow}^sN_{\uparrow,\downarrow}^f(1-\epsilon)]\
\\&-[N_{\downarrow,\uparrow}^sN_{\downarrow,\uparrow}^f+N_{\uparrow,\downarrow}^sN_{\uparrow,\downarrow}^f]\
\\
&= \frac{-\epsilon N^s N_{\uparrow,\downarrow}^f}{2}.
\end{split}
\end{align}
 From this the polarization follows,
\be
\label{Polarization}
P=\left|\frac{N^f_{\uparrow}
-N^f_{\downarrow}}{N^f_{\uparrow}+N^f_{\downarrow}}\right|
=\left|\frac{\delta_{+} - \delta_{-}}{\delta_{+} + \delta_{-}}\right|.
\ee

\begin{figure}
\begin{flushleft}
\includegraphics[width=.48\textwidth]{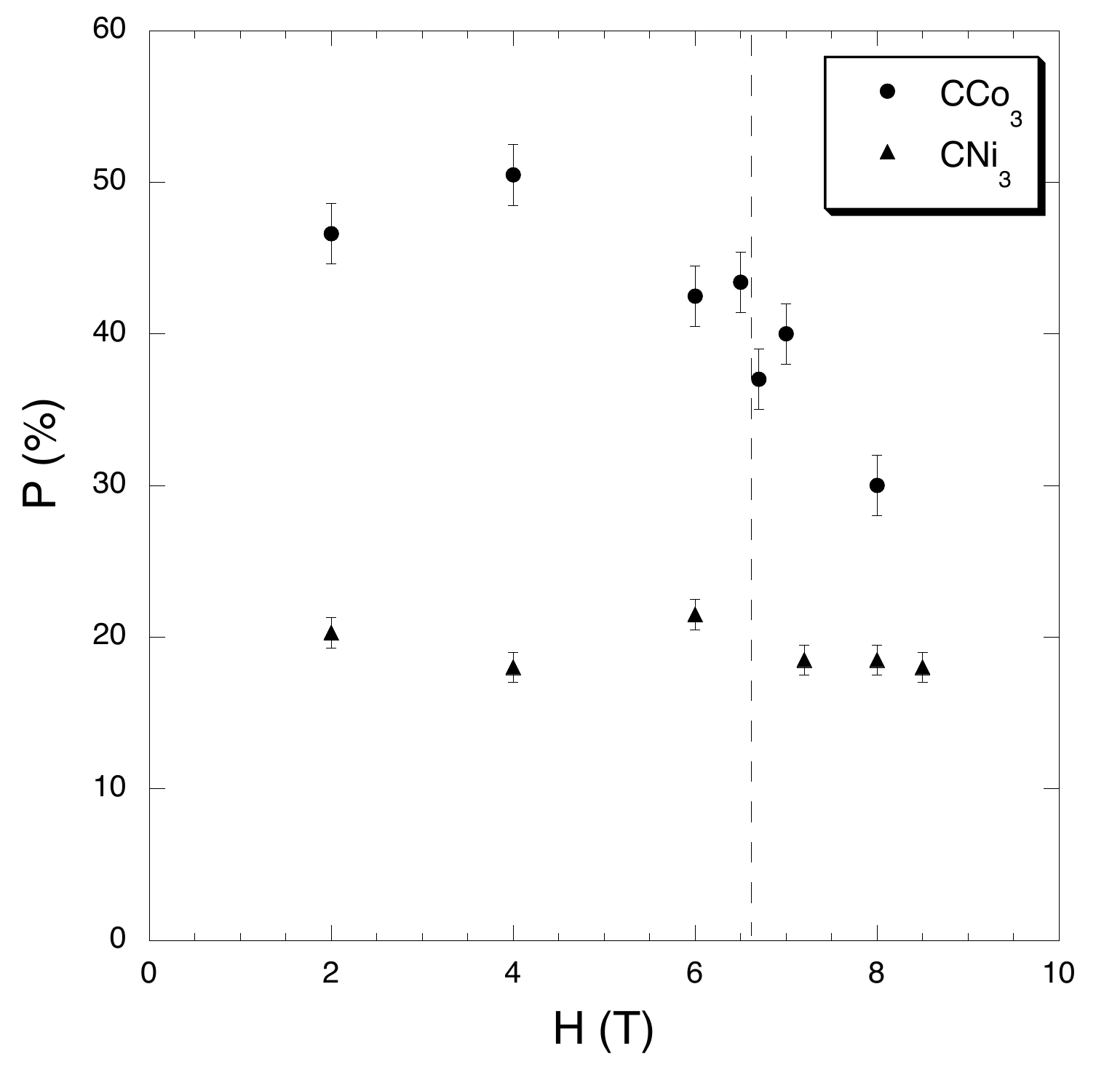}\end{flushleft}
\caption{\label{P-H}Electron polarization of 45~\AA\ CNi$_3$ and
CCo$_3$ films as a function of magnetic field.  The dashed line
represents the approximate parallel critical field of the Al
counter-electrodes used in these measurements.}
\end{figure}

The data in Fig.~\ref{PR} give a polarization of 39.5\% which is
comparable to values reported for pure Co by the early work of
Tedrow and Meservey \cite{TM2} and later measurements using PCAR
\cite{Soulen1998}.  We note that \eref{Polarization} is independent
of both the PR's strength $\epsilon$ and its width.  Thus within the
limits of signal-to-noise constraints, one should be able to obtain
polarization measurements at quite high fields.  The dashed line in
Fig.~\ref{PR} represents a prediction for the resonance curve at
18~T obtained by extrapolating the necessary parameters from fitting
the data at 6.7~T.
Although the PR is significantly attenuated and broadened by the
high field, it is still well within the noise level of the data.
Since the strength of the resonance grows as
the conductance $g$ is reduced, one can use slightly thinner
counter-electrodes to increase the visibility of the resonance.
Previous studies have shown that the PR in counter-electrodes with
$g\le6$ remains well defined in perpendicular fields of a few Tesla,
thus eliminating the need for precise parallel alignment
\cite{Wu2005}.

In Fig.~\ref{P-H} we present electron polarization as a function of
magnetic field for CNi$_3$ and CCo$_3$ films.  The vertical dashed
line represents the approximate critical field of the
counter-electrodes used in this study.  The data show good
agreement between polarization values measured just below $H_{c||}$
and normal-state values measured just above $H_{c||}$.  The
polarization of CNi$_3$ film is independent of field, as would be
expected considering the fact that the Zeeman energies associated
with fields of a few Tesla are less than 0.1~meV, which is much less
than exchange energies associated with the
magnetization.  Interestingly, though, the polarization of the CCo$_3$
film exhibits a significant decrease in fields above 3-4~T.  This
trend can be seen in both the superconducting phase measurements and
the PR measurements.  The origin of this field dependence is
uncertain.

In summary, we have shown that conduction spin polarization can be
determined from the relative amplitudes of the occupied and
unoccupied pairing resonance features.  The technique can be used in
fields that are several times higher than the Clogston-Chandrasehkar
critical field, thus allowing polarization measurements to be made over
a very wide range of magnetic fields.  Preliminary polarization
measurements in CCo$_3$ films show a strong suppression of the
polarization in fields above a few Tesla.

\acknowledgments

We gratefully acknowledge enlightening discussions with Ilya
Vekhter
and Dan Sheehy. This work was supported by the DOE under
Grant No.\
DE-FG02-07ER46420.


\begin{thebibliography}{30}
\bibitem{Wolf2001} S. A. Wolf, D. D. Awschalom, R. A. Buhrman, J. M.
Daughton, S. von Moln½r, M. L. Roukes, A. Y. Chtchelkanova, and D. M.
Treger, Science {\bf 294}, 1488 (2001).
\bibitem{Prinz1995} G. Prinz, Phys. Today {\bf 48}, 58 (1995).
\bibitem{TM1} P.M. Tedrow and R. Meservey, Phys. Rev. Lett. {\bf 26},
192 (1971).
\bibitem{TM2} P.M. Tedrow and R. Meservey, Phys. Rep. {\bf 238}, 173 (1994).
\bibitem{Johnson1997} P.D. Johnson, Rep. Prog. Phys. {\bf 60}, 1217 (1997).
\bibitem{Buhrman1998} S.K. Upadhyay, A. Palanisami, R.N. Louie, and
R.A. Buhrman, Phys. Rev. Lett. {\bf 81}, 3247 (1998).
\bibitem{Soulen1998} J. Soulen, Jr., J. M. Byers, M. S. Osofsky, B.
Nadgorny, T. Ambrose, S. F. Cheng, P. R. Broussard, C. T. Tanaka, J.
Nowak, J. S. Moodera, A. Barry, and J. M. D. Coey, Science {\bf 282},
85 (1998).
\bibitem{Buhrman2007} P. Chalsani, S.K. Upadhyay, O. Ozatay, and R.A.
Buhrman, Phys. Rev. B {\bf 75}, 094417 (2007).
\bibitem{Aleiner1997}I. L. Aleiner and B. L. Altshuler, Phys. Rev.
Lett. {\bf 79}, 4242 (1997).
\bibitem{Butko1999}V. Y. Butko, P. W. Adams, and I. L. Aleiner, Phys.
Rev. Lett. {\bf 82}, 4284 (1999);
H. Y. Kee, I. L. Aleiner, and B. L. Altshuler, Phys. Rev. B {\bf 58},
5757 (1998).
\bibitem{Fulde1973}P. Fulde, Adv. Phys. {\bf 22}, 667 (1973).
\bibitem{TM3} R. Meservey, P.M. Tedrow, and P. Fulde, Phys. Rev.
Lett. {\bf 25}, 1270 (1970).
\bibitem{CNi3} D.P. Young, A.B. Karki,
P.W. Adams, J.N. Ngunjiri, J.C. Garno, H. Zhu, B. Wei, D. Moldovan,
J. App. Phys. {\bf 103}, 053503 (2008).
\bibitem{Clogston1962}A. M. Clogston, Phys. Rev. Lett. {\bf 9}, 266 (1962).
\bibitem{Chandrasekhar1962}B. S. Chandrasekhar, Appl. Phys. Lett.
{\bf 1}, 7 (1962).
\bibitem{CWA}G. Catelani, X. S. Wu, and P. W. Adams, Phys. Rev. B
{\bf 78}, 104515 (2008).
\bibitem{Altshuler1987} B.L. Altshuler, A.G. Aronov, M.E. Gershenson, and Yu.V.
Sharvin, Sov. Sci. Rev. A. Phys. Vol. {\bf9}, 223 (1987).
\bibitem{Imry1982} Y. Imry and Z. Ovadyahu, Phys. Rev. Lett.
{\bf49}, 841 (1982).
\bibitem{Catelani2009} G. Catelani, Y. M. Xiong, X. S. Wu, P. W.
Adams, e-print \href{http://arxiv.org/abs/0905.2414}{arXiv:0905.2414}.
\bibitem{Wu2005} X.S. Wu, P.W.
Adams, and G. Catelani, Phys. Rev. Lett. {\bf95}, 167001 (2005).
\end{thebibliography}
\end{document}